# DIGITAL RESILIENCE AND THE CONTINUANCE USE OF MOBILE PAYMENT SERVICES

Muftawu Dzang Alhassan, Stellenbosch University, South Africa, 25727907@sun.ac.za

Martin Butler, Stellenbosch University, South Africa

**Abstract:** The use of mobile payment services is an essential contributor to financial inclusion in emerging markets. Unfortunately, the service has become a platform for fraud. Mobile payment users need to be digitally resilient to continue using the service after adverse events. However, there is scant literature on users' continuance use of mobile payment services in the post-event of fraud. The focal point of prior literature has been on technology adoption or threat avoidance to implement policies that protect users. Analysing the relationship between individual digital resilience and post-adoption behavioural patterns will enable service providers to support individual digital resilience to promote users' continuance use of the service. This research aims to develop and empirically validate a conceptual model to examine individual digital resilience in the context of the continuance use of mobile payments. The model will be based on protection motivation theory. Survey data will be obtained from victims of mobile payment fraud and other users who continue using the service despite their knowledge of mobile payment fraud. The results from this study are expected to make key contributions to theory, practice, and policy in the areas of digital resilience, mobile payments, and ICT4D.

## 1. INTRODUCTION

The growth of the mobile payment services sector is an essential contributor to financial inclusion in developing countries (Senyo & Osabutey, 2020). This sector has facilitated individuals' access to low-cost and reliable financial services, especially in developing countries (Liébana-Cabanillas et al., 2019).

Mobile payment service innovations such as m-pesa in Kenya and Tanzania have transformed into ubiquitous mobile payment platforms (Varga, 2017) that enable users to move funds at their convenience (Iman, 2018). The mobile payment success was replicated in Ghana by the Mobile Telecommunication Network (MTN) in 2009. This was followed by other telecommunication providers Vodafone and AirtelTigo, leading to increased mobile payment transactions and active users (Alhassan et al., 2020).

Unfortunately, mobile payment services have become an avenue for fraud (Akomea-Frimpong et al., 2019). This is particularly evident in developing countries where high levels of fraud are apparent (Provencal, 2017). Cybercriminals often rely on social engineering approaches to lure mobile money subscribers into revealing their mobile credentials that can be used to illegally withdraw money from the users' mobile money wallets (Annan, 2017; Pradigdya et al., 2019). More sophisticated attacks involve taking control over a user's mobile device to access payment applications or telecommunications systems (Ali et al., 2019).

Service providers should pay more attention to the safety of users' information and financial assets due to widespread issues of fraud in mobile payments (Humbani & Wiese, 2020; Ofori et al., 2017). When mobile payments are secure and financial service providers guarantee users' information and funds against fraudsters, users' trust in the service will increase, leading to the continuance use of the service (Kumar et al., 2017; Odoom & Kosiba, 2020).

Despite several measures instituted by mobile payment service providers to help curb fraud (Priezkalns, 2020), an increasing number of users report being victims of mobile payment fraud (Ankiilu, 2017). A total of 365 mobile money fraud cases are reported daily by MTN users in Ghana (Larnyoh, 2020). In spite of fraudulent events, the use of mobile payments continues to grow. Significant research is directed at mobile technology adoption and threat avoidance (Butler, 2020), literature exploring the intrinsic factors that ensure individuals' continuance use of technologies remains limited (Liao et al., 2009).

The study proposes *'individual digital resilience'* as a capability that enable users to continue using mobile payments in the post-event of fraud. Our research aims to answer the following question: *What are the antecedents of individual digital resilience and the effect thereof of the continuance of mobile payment services?*

## 2. PRIOR LITERATURE

### 2.1. Digital Resilience

The concept of resilience emerged from ecological studies of the 1970s. Holling (1973) introduced resilience and proposed that "resilience determines the persistence of relationships within a system and is a measure of the ability of these systems to absorb changes of state variables, driving variables, and parameters and still persist" (Holling, 1973, p.18). When systems break down as a result of disruptions, they must recover, preferably, to their former state (Zhang & Zhao, 2019). Recovery will ensure business continuity and mitigate losses arising from adverse events (Onwubiko, 2020).

In recent years, the term resilience has been applied in several contexts and for different concepts. In Information Systems (IS) research, the term *digital resilience* has emanated and gained popularity due to the emergence of digital and cyber-attacks that threaten individuals, companies, and governments (Hammond & Cooper, 2015; Kohn, 2020a). Organisational (digital) resilience is actively researched and can be described as recovery from digital attacks on data, networks, and computers (Rothrock, 2018).

However, resilience also refers to individual capabilities. Literature at the individual level primarily deals with digital skills and capabilities (van Laar et al., 2020) and less with individual digital resilience at the same level as organisational resilience. Individual resilience is presented in IS literature as the ability to avoid or withstand a cyberattack or the degree of information security awareness (McCormac et al., 2018; Udwan et al., 2020). In the omnipresence of cyber threats, and given the initial conceptualisation of the construct of individual resilience, it is also essential for users of technology to develop resilience to recover from adverse cyber events.

For this study, we define individual digital resilience as the ability of a user to withstand and recover from an adverse technology usage event. The study will use mobile payment services to investigate the phenomena due to the widespread use despite the occurrence of fraudulent events. Prior studies at the individual level have related resilience to traits such as threat appraisal, coping (O'Leary, 1998), perseverance, positive emotion, meaning-making, and growth (Amir, 2012; Connor & Davidson, 2003; Wagnild & Young, 1993).

Digital resilience studies at the individual level have primarily examined how individuals can leverage technologies to build resilience to prevent disruptions (Tim et al., 2020; Udwan et al., 2020). Users' digital resilience in the event of disruptions and its effect on the continuance use of the service or technology has received less attention. Literature (Camp et al., 2019; Majchrzak et al., 2018) has called for studies to examine individual digital resilience from different perspectives. For instance, Kohn (2020b) recommended studies to explore network effects of resilience on individuals and work communities. Liao et al. (2009) introduced the Technology Continuance Theory (CTC) whilst acknowledging that multiple factors, like expectation confirmation, requires further scrutiny to understand the impact on the continuance of technology usage.

Understanding individual digital resilience and how it influences post behavioural intent will enable technology developers, service providers, and governments to prioritise their limited resources and promote intended benefits to protect users' information and funds.

## 2.2. Digital Resilience in Mobile Payments

Digital financial services can reduce transaction costs and extend an individual's access to greater social connections (Lyons et al., 2019). In the event of disruptions, individuals can rely on these social connections that contribute to their resilience because of the systemic effect of the networks (Jack & Suri, 2014). Being digitally resilient is more than a skillset to cope with using technologies.

Krishnan, Johri, Chandrasekaran and Pal (2019) argued that the adoption of digital payments enabled Indian citizens to build resilience when the government decided to end the use of a particular legal tender. The study showed that individuals who possessed adequate digital skills were more resilient to the disruption and continued to rely on digital payments for cash transactions. In his instance, poor individuals in rural settlements exhibited greater resilience than poor urban dwellers showing the complexity of the attributes that could potentially define and contribute towards digital resilience.

Fraud has become a universal threat that negatively affects the survival of individuals, businesses, and economies (Kovács & David, 2016). Mobile payment services are no exception to this vexing problem (Akomea-Frimpong et al., 2019), with mobile money fraud in Ghana increasing from 278 cases in 2015, to 388 in 2016 (Ankiilu, 2017).

| Indicators | 2015 | 2016 | 2017 | 2018 |
|---|---|---|---|---|
| Total number of registered mobile money accounts | 13,120,367 | 19,735,098 | 23,947,437 | 32,554,346 |
| Active mobile money accounts | 4,868,569 | 8,313,283 | 11,119,376 | 13,056,978 |

**Source:** Bank of Ghana (2018)

**Table 1. Mobile payment usage in Ghana**

Table 1 indicates the growth in mobile payment usage in Ghana from the year 2012 to 2018. It is evident that individuals continue to use the service, despite issues of reported fraud (Yeboah, 2021).

## 2.3. Antecedents of digital resilience

Research in technology adaption has evolved significantly with the multiple versions of technology acceptance models (Tsai et al., 2016) and threat avoidance models prominent in the IS literature. The antecedents of technology acceptance, like performance expectancy, effort expectancy, facilitating conditions, social influence, and perceived security, are actively researched (Legowo, 2019).

Conversely, factors that lead to higher levels of individual resilience are not part of the active academic discourse, nor have they been presented concisely. Still, the development of digital skills is widely acknowledged as a driver of digital resilience. Van Laar (2020) noticed a lack of appreciation of social determinants such as social support in understanding digital skills and argued for a comprehensive view on the skills dimension. Digital resilience potentially encapsulates other factors like the facilitating conditions, risk propensity, perceived usefulness, trust and self-efficacy (Chang, 2010; Susanto et al., 2016). As a construct, there are many potential drivers of individual digital resilience, but this is not yet fully defined or widely used.

We intend to address both the lack in definition and measurement of digital resilience and the impact of digital resilience on the continuance of the use of technology. Using a service like mobile payments that 1) continues to grow, 2) seems resilient in its operations and 3) is subject to adverse events provides a relevant context to explore the phenomena. Exploring the factors that contribute to higher levels of individual digital resilience is essential in an increasingly digitised world.

## 3. PROPOSED THEORY

To answer the research question, a conceptual model for understanding individual digital resilience based on a modification of the Protection Motivation Theory (PMT) from Rogers (1975) will be created.

The basic tenets of the PMT are that when confronted with a threat(s), an individual experiences two cognitive processes of threat appraisal and coping appraisal. *Threat appraisal* involves a process of analysing (1) perceived threat vulnerability, and (2) perceived threat severity. *Coping appraisal,* on the other hand, involves evaluating (1) the efficacy of the potential adaptive responses to a threat (*response efficacy)*; (2) the ability to successfully carry out the recommended responses (*self-efficacy*); and (3) the *response costs* associated with the engagement in an adaptive coping strategy.

The PMT has been adapted and applied in different contexts. In IS research, for example, the PMT has been used to investigate information security behaviours (Hassandoust & Techatassanasoontorn, 2018; Yang et al., 2020). Studies in this area show how users engage in secure behaviours to protect their data and information from intruders (Giwah et al., 2019). However, while the PMT has been able to significantly explain individuals' secure behaviour, there is proof that extended versions of the PMT are more accurate under certain conditions (Aurigemma & Mattson, 2018; Ifinedo, 2012).

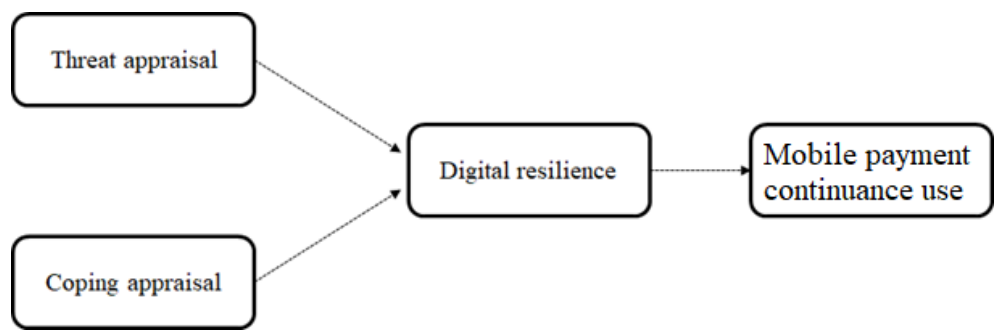

**Figure 1. Proposed research model**

The extension proposed in this study is done by including *individual digital resilience* as a construct that is influenced by the threat and coping appraisal dimensions of the PMT. This is indicated in Figure 1. The rationale for relying on the PMT in our study is that past studies highlighted threat appraisals and coping appraisal (such as self-efficacy) as dimensions of individual resilience (O'Leary, 1998; Ledesma, 2014).

The use of PMT and individual digital resilience will enable the study to examine how (1) threat and coping appraisal dimensions of PMT influence individual digital resilience, and (2) individual digital resilience influences the continuance use of mobile payment services.

# 4. PROPOSED METHODOLOGY

A survey instrument will be developed to collect data from users of mobile payment services that have been victims of fraud or knows about fraud and continue using mobile payment services. Data will also be collected from mobile phone users that have stopped using mobile payment services. The qualifying question for inclusion is thus not current usage but prior or current usage of mobile payment services.

Measurement items will be adapted from previous studies and measured using a five-point Likert scale. *Individual digital resilience* will be adapted from the studies by Amir (2012) and include additional items that define antecedents of digital resilience from a comprehensive literature review. The measurement items for *threat* and *coping appraisal* will use questions developed by Liang and Xue (2010) and Tsai, Jiang, Alhabash, Larose, Rifon and Cotten (2016). Measurement items for the *continuance use of mobile payment services* will be drawn from Shao, Zhang and Guo (2019).

Additional measures that could influence individual digital resilience from studies that used and adapted the *Technology Continuance Theory* (Liao et al., 2009)will be added to the model to ensure a rich data set for further analysis. Extensive demographic data, including education, geographical location, and other factors that could play a role in risk perception, will be included. A measure of digital skills and social influences will also be included in the research instrument.

The survey will be administered in Ghana due to the richness of the available information in a country with high mobile payment usage and regular fraudulent events to whom users may have been exposed or have knowledge of. Getting users who decided to continue and discontinue mobile payment services are essential for the study to determine the model's accuracy.

We will analyse the survey data using Partial Least Squares-Structural Equation Modelling (PLS-SEM). PLS-SEM will enable the researcher(s) to examine the effect sizes between the variables in the model. The researchers will investigate the mediating effects of individual digital resilience on the linkages between the threat and coping appraisal dimensions of the PMT and mobile payment continuance use. Hair, Risher, Sarstedt and Ringle (2019) argued that PLS-SEM was appropriate for studies that tested models with second-order constructs and mediating variables.

# 5. EXPECTED RESULTS

The objective of the research is to examine the antecedents of individual digital resilience on the continuance of mobile payment services use. Multiple outcomes are expected.

First, we expect to empirically confirm that mobile payment users will engage in threat and coping appraisals before and after falling victim to fraud, or gaining knowledge about fraudulent activities. Through this, they may be growing their level of digital resilience to continue to use the service, irrespective of adverse events.

However, the extent of digital resilience may vary amongst users of mobile payment services. Some users may display high levels of resilience to continue using the service whilst others may not. It is anticipated that users with lower levels of digital resilience will show a higher propensity to discontinue mobile payment services after experiencing an adverse event or learning about these events.

Furthermore, the study will examine the mediating effects of individual digital resilience on the nexus between the threat and coping appraisal dimensions of the PMT and mobile payment continuance use.

Finally, other constructs that may influence individual digital resilience from existing theoretical models will be analysed to potentially enhance the structural model to improve the ability to predict the continuance of use and provide new directions for research on individual digital resilience.

# 6. CONCLUSION AND EXPECTED CONTRIBUTION

Results from this study are expected to make several contributions to theory and practice in the areas of digital resilience, mobile payments, and ICT4D.

First, this study is novel in the combination of the PMT and digital resilience to examine individual digital resilience and the continuance use of mobile payment services in a developing country. This combination will provide valuable insights into how (1) mobile payment users develop resilience through threat and coping appraisals and (2) individual digital resilience influence the continuance use of the service in the post-event of fraud.

Second, the study contributes to the limited research on individual digital resilience with previous literature focusing on organisational, community, and technology infrastructure digital resilience. The development and validation of an instrument to measure digital resilience, based on robust theory, will contribute to research options in this underdeveloped area of growing interest.

Practitioners and policymakers will be able to use the research results to develop interventions that will develop individual digital resilience. The study will enhance the continuance use of mobile payment services by defining appropriate interventions.

# REFERENCES AND CITATIONS